\documentclass[a4paper]{article}

\usepackage[dvipsnames]{xcolor}  
\definecolor{titleblue}{rgb}{0.16,0.24,0.64} 
\definecolor{citecolor}{rgb}{0.2,0.3,0.8}

\usepackage{graphicx} 
\usepackage{mathtools}

\usepackage{hyperref}

\hypersetup{
    colorlinks    = true, 
    citecolor    = citecolor, 
    linkcolor    = red, 
    urlcolor    = magenta, 
}

\usepackage{breakurl}
\usepackage{stmaryrd}
\usepackage{wasysym}

\usepackage{multirow}

\makeatletter
\def\verbatim{\small\@verbatim \frenchspacing\@vobeyspaces \@xverbatim}
\makeatother



\begin{document}


\title{Lup-Like Cantilever Beam for Small Deflection}
\author{Sparisoma Viridi\thanks{viridi@fkt.physik.tu-dortmund.de} \\[0.1cm]
Department of Physic, Faculty of Mathematics and Natural Sciences, \\
Institut Teknologi Bandung, Bandung 40132, Indonesia \\
Lehrstuhl f\"ur Theoretische Physik I, Fakult\"at Physik, \\
Technische Unversit\"at Dortmund, D-44227 Dortmund, Germany \\[0.2cm]
Mitra Djamal \\[0.1cm]
Department of Physic, Faculty of Mathematics and Natural Sciences, \\ Institut Teknologi Bandung, Bandung 40132, Indonesia \\[0.2cm]
Yusaku Fujii \\[0.1cm]
Department of Electronic Engineering, Faculty of Engineering, \\
Gunma University, Kiryu 376-8515, Japan}
\date{23 February 2014}
\maketitle

\begin{abstract}
A lup-like cantilever beam are discussed in this work. For small deflection it can be approximated as a spring-mass system with certain spring constant whose effective mass is larger than the usual constant rectangular cross section cantilever beam. A new parameter $\beta$ is introduced to relates some the properties of lup-like cantilever beam to the usual one. Influence of beam witdh $B_0$ and head width $B_t$ to value of $\beta$ is also presented.
\end{abstract}

\section{Introduction}

Cantilever beams play important role in many today applications. It is used as components in common bridge \cite{Shama_2001}, railway bridge \cite{Koyama_1997}, and aeroplane wing \cite{Armanios_1995}. In smaller scale it is in sensors for viscosity \cite{Bergaud_2000}  and acceleration \cite{Fricke_1993}. In nanoscopic scale application for atomic force microscope (AFM) is already common \cite{Sader_1998}, even it can be used to measure weight of single virus \cite{Gupta_2004}. Common form for cantilever beam is with constant rectangular cross section, where different form will have its own first mode natural frequency \cite{Hoffmann_2000}. A recent application uses also a lup-like form which is not yet common \cite{2014arXiv1402.2153D}, that needs a theoretical approach to characterize the cantilever beam, which is discussed in this work. Limitation to small deflection is still required here.

\section{Mass and area of moment inertia}

\begin{figure}
\center
\includegraphics[width=10cm]{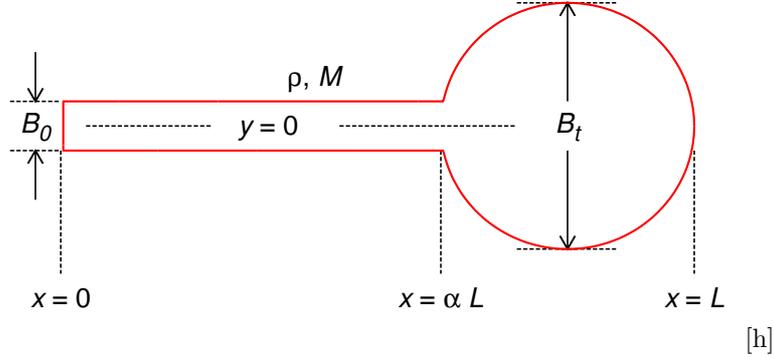}[h]
\caption{Model of lup-like cantilever beam with mass $M$, density $\rho$, arm width $B_0$, arm length $\alpha L$, head width $B_t$, head length $B_t - \delta$, and thickness $H$ (in $z$ direction, perpendicular to drawing plane).}
\label{fig:lup-like-cantilever-beam-xy}
\end{figure}

A cantilever beam that has a lup-like form is illustrated in Figure \ref{fig:lup-like-cantilever-beam-xy}. It has length of $L$, thickness $H$, density $\rho$, and mass $M$. The cantilever consists of two parts, which are arm and head. Arm has length of $\alpha L$ and width of $B_0$, while head has length of $B_t - \delta$ and width of $B_t$.

A function to represent width of cantilever beam $B$ as function of $x$ for this case can be defined as

\begin{equation}
\label{eqn:cantilever-width}
B(x) = \left\{
\begin{array}{lr}
B_0, & 0 \le x < \alpha L, \\
&\\
2\sqrt{\frac14 B_t^2 - \left[x - \left(\alpha L + \frac12 \sqrt{B_t^2 - B_0^2}\right)\right]^2}, & \alpha L \le x \le L,
\end{array}
\right.
\end{equation}

with head length to beam length ratio $(1 - \alpha)$ defined as

\begin{equation}
\label{eqn:one-minus-alpha}
(1 - \alpha) = \frac{B_t - \delta}{L},
\end{equation}

and parameter $\delta$ as

\begin{equation}
\label{eqn:delta}
\delta = \frac12 \left( B_t - \sqrt{B_t^2 - B_0^2} \right),
\end{equation}

which makes the cross section of the beam

\begin{equation}
\label{eqn:cantilever-cross-section}
A(x) = \left\{
\begin{array}{lr}
H B_0, & 0 \le x < \alpha L, \\
&\\
2H\sqrt{\frac14 B_t^2 - \left[x - \left(\alpha L + \frac12 \sqrt{B_t^2 - B_0^2}\right)\right]^2}, & \alpha L \le x \le L.
\end{array}
\right.
\end{equation}

Mass of the beam with constant density $\rho$ is then determined using

\begin{equation}
\label{eqn:mass-rho-integral}
M = \int_0^L \rho A(x) dx,
\end{equation}

which gives result

\begin{equation}
\label{eqn:cantilever-mass}
M = \alpha \rho H B_0 L + \frac14 (1 - \alpha) \rho H B_t L \left[ \pi - sin^{-1} \left(\frac{B_0}{B_t}\right) + \frac{B_0\sqrt{B_t^2 - B_0^2}}{B_t^2} \right].
\end{equation}

In Equation (\ref{eqn:cantilever-mass}) parameter $\alpha$ and $B_t$ are dependent to each other, e.g. value $\alpha = 1$ corresponds to $B_t = 0$, while $\alpha = 0$ corresponds to $B_t = L$ as in Equation (\ref{eqn:one-minus-alpha}), and at these limits it is required that $\delta = 0$.

Area of moment inertia for a cantilever beam with width $B$ which is deflected in the direction of its thickness $H$ is \cite{Rabe_1996}

\begin{equation}
\label{eqn:cantilever-area-of-moment-inertia}
I = \frac{1}{12} H^3 B.
\end{equation}

In this work width of the beam is not constant but function of $x$ as it is previously given in Equation (\ref{eqn:cantilever-width}), then it turns Equation (\ref{eqn:cantilever-area-of-moment-inertia}) into

\begin{equation}
\label{eqn:cantilever-area-of-moment-inertia-x}
I(x) = \frac{1}{12} H^3 B(x).
\end{equation}

\section{Small deflection}

\begin{figure}[h]
\center
\includegraphics[width=10cm]{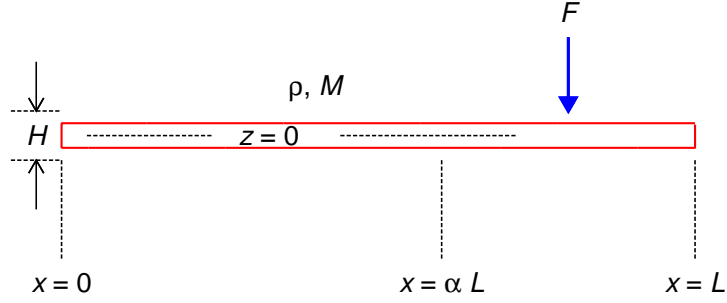}
\caption{A force $F$ is applied to center of the head of lup-like cantilever beam.}
\label{fig:lup-like-cantilever-beam-zx}
\end{figure}

The lup-like cantilever beam is tipped in the center of the head in $z$ direction (perpendicular to the drawing plane in Figure \ref{fig:lup-like-cantilever-beam-xy} or along the drawing plane in Figure \ref{fig:lup-like-cantilever-beam-zx}) so it deflects. For linear analysis of small deflection, the curvature $\kappa$ of the deflected beam is approximated as \cite{Zhang_2004}

\begin{equation}
\label{eqn:small-deflection-curvature}
\kappa = \frac{d^2z}{dx^2}.
\end{equation}

There is relation between curvature $\kappa$, elastic modulus $E$, bending moment along $x$ axis $\tau(x)$, and area of moment inertia $I$ \cite{Bucciarelli_2002}

\begin{equation}
\label{eqn:curvature-bending-moment-e}
\kappa = \frac{\tau(x)}{EI}.
\end{equation}

If a force $F$ is applied to the center of cantilever head as illustrated in Figure \ref{fig:lup-like-cantilever-beam-zx} with fixed end at $x = 0$ and free end at $x = L$, then the bending moment would be

\begin{equation}
\label{eqn:bending-moment-x}
\tau(x) = \left[ x - \left( L - \frac12 B_t \right) \right] F.
\end{equation}

Substitute Equations (\ref{eqn:small-deflection-curvature}) and (\ref{eqn:bending-moment-x}) into Equation (\ref{eqn:curvature-bending-moment-e}) will produce a second order differential equation

\begin{equation}
\label{eqn:differential-equation-I-constant}
\frac{d^2z}{dx^2} = \frac{F}{E I} ~ \left[ x - \left( L - \frac12 B_t \right) \right], \end{equation}

whose solution is

\begin{equation}
\label{eqn:deflection-x-I-constant}
z(x) = \frac{F}{EI} \left[ \frac16 x^3 - \frac12 \left( L - \frac12 B_t \right) x^2 \right]
\end{equation}

which is a little bit different than for uniform rectanguler cross section cantilever beam, which is tipped in the free end of the beam \cite{Whitney_1999}. Equation (\ref{eqn:deflection-x-I-constant}) also assumes that area of moment inertia is constant.

Different result will be obtained if Equation (\ref{eqn:cantilever-area-of-moment-inertia-x}) is substituted first into Equation (\ref{eqn:curvature-bending-moment-e}) before solving the second order differential equation. Then following second order differential equation will be produced

\begin{eqnarray}
\nonumber
\frac{d^2z}{dx^2} = \frac{12 F}{E H^3} ~ \left[ x - \left( L - \frac12 B_t \right) \right] \times \\
\nonumber
\\
\label{eqn:differential-equation-I-not-constant}
\left\{
\begin{array}{lr}
B_0^{-1}, & 0 \le x < \alpha L, \\
&\\
\left(2\sqrt{\frac14 B_t^2 - \left[x - \left(\alpha L + \frac12 \sqrt{B_t^2 - B_0^2}\right)\right]^2}\right)^{-1}, & \alpha L \le x \le L.
\end{array}
\right.
\end{eqnarray}

For $0 \le x < \alpha L$ the solution of Equation (\ref{eqn:differential-equation-I-not-constant}) is similar to Equation (\ref{eqn:deflection-x-I-constant}), which is

\begin{equation}
\label{eqn:deflection-x-I-not-constant-0}
z(x) = \frac{12F}{E H^3 B_0} \left[ \frac16 x^3 - \frac12 \left( L - \frac12 B_t \right) x^2 \right].
\end{equation}

And for $\alpha L \le x \le L$, following constants

\begin{eqnarray}
\label{eqn:c1}
c_1 = \frac{6F}{EH^3}, \\
\label{eqn:c2}
c_2 = L - \frac12 B_t = \alpha L + \frac12 \sqrt{B_t^2 - B_0^2}, \\
\label{eqn:c3}
c_3 = \frac12 B_t,
\end{eqnarray}

and also functions and other constants

\begin{eqnarray}
\label{eqn:cos-theta}
\cos \theta(x) = \frac{x - c_2}{c_3}, \\
\label{eqn:sin-theta}
\sin \theta(x) = \frac{\sqrt{c_3^2 - (x - c_2)^2}}{c_3}, \\
\label{eqn:cos-theta-alpha}
\cos \theta_\alpha =  \frac{\sqrt{B_t^2 - B_0^2}}{B_t^2}, \\
\label{eqn:sin-theta-alpha}
\sin \theta_\alpha = \frac{B_0}{B_t}.
\end{eqnarray}

are defined. Constants and functions in Equation (\ref{eqn:c1}) - (\ref{eqn:sin-theta-alpha}) will simplify Equation (\ref{eqn:differential-equation-I-not-constant}) for  $\alpha L \le x \le L$ to

\begin{equation}
\label{eqn:differential-equation-I-not-constant-L}
\frac{d^2z}{dx^2} = \frac{c_1 (x - c_2)}{ \sqrt{c_3^2 - (x - c_2)^2} }
\end{equation}

First integration from $x = \alpha L$ to $x$ will turn Equation (\ref{eqn:differential-equation-I-not-constant-L}) into

\begin{equation}
\label{eqn:deflection-dzdx}
\frac{dz}{dx} - \left. \frac{dz}{dx} \right|_{x = \alpha L} = c_1 \sqrt{c_3^2 - (x - c_2)^2} - \frac12 c_1 B_0.
\end{equation}

Further integration within the same range will lead to

\begin{eqnarray}
\label{eqn:deflection-zx}
z(x) - z(\alpha L) = \frac{c_1 c_3^2}{2} \left[ \sin \theta(x) \cos \theta(x) - \frac{B_0 \sqrt{B_t^2 - B_0^2}}{B_t^2} \right. \\
\left. - \theta(x) + \sin^{-1} \left(\frac{B_0}{B_t}\right) \right] + \left[ \left. \frac{dz}{dx} \right|_{x = \alpha L} - \frac{c_1 B_0}{2} \right] (x - \alpha L)
\end{eqnarray}

Values of $z(\alpha L)$ and its derivative $[dz/dx](\alpha L)$ are obtained from Equation (\ref{eqn:deflection-x-I-not-constant-0}), which are

\begin{eqnarray}
\label{eqn:deflection-boundary-z}
z(\alpha L) = \frac{12 F}{E H^3 B_0} \left[ \frac12 \alpha^2 L^3 \left( \frac{\alpha}{3} - 1 \right) + \frac14 \alpha^2 L^2 B_t \right], \\
\label{eqn:deflection-boundary-dzdx}
\left. \frac{dz}{dx} \right|_{x = \alpha L} = \frac{12 F}{E H^3 B_0} \left[ \alpha L^2 \left( \frac{\alpha}{2} - 1 \right) + \frac12 \alpha L B_t \right].
\end{eqnarray}

Using Equations (\ref{eqn:deflection-boundary-z}) and (\ref{eqn:deflection-boundary-dzdx}) and by setting $x = c_2$ deflection of the head of cantilever beam where the force $F$ tips it can be found, which is

\begin{eqnarray}
\nonumber
z(c_2) = \frac{12 F}{E H^3 B_0} \left[ \frac12 \alpha^2 L^3 \left( \frac{\alpha}{3} - 1 \right) + \frac14 \alpha^2 L^2 B_t \right] \\
\nonumber
- \frac{3 F B_t^2}{4 E H^3} \left[ \frac{B_0 \sqrt{B_t^2 - B_0^2}}{B_t^2} + \frac{\pi}{2} - \sin^{-1} \left(\frac{B_0}{B_t}\right) \right] \\
\label{eqn:deflection-x-I-not-constant-1}
+ \frac12 \sqrt{B_t^2 - B_0^2} \left\{ \frac{12 F}{E H^3 B_0} \left[ \alpha L^2 \left( \frac{\alpha}{2} - 1 \right) + \frac12 \alpha L B_t \right] - \frac{3 F B_0}{E H^3} \right\}.
\end{eqnarray}

Identity from Equation (\ref{eqn:c2}) can simplify Equation (\ref{eqn:deflection-x-I-not-constant-1}) into

\begin{eqnarray}
\nonumber
z(c_2) = \frac{F}{E H^3} \left\{ \frac{9}{2} B_0 \left[ (1 - \alpha) L + \frac12 B_t \right] + \frac34 B_t^2 \left[ \sin^{-1} \left( \frac{B_0}{B_t} \right) - \frac{\pi}{2} \right] \right. \\
\label{eqn:deflection-x-I-not-constant-2}
\left. + \frac{6 \alpha L B_t}{B_0} \left[ (2 - \alpha)L - \frac12 B_t \right] - \frac{12 L^3}{B_0} \left( \alpha - \alpha^2 + \frac13 \alpha^3 \right) \right\}.
\end{eqnarray}

For $\alpha = 1$ and $B_t = 0$ Equations (\ref{eqn:deflection-x-I-not-constant-2}) and (\ref{eqn:deflection-x-I-constant}) give the same result, which can be considered as proof for the first equation.

\section{Spring constant and natural frequency}

For small deflection in $z$ direction of a cantilever beam with one fixed end and the other end is under influence of certain force $F$, the beam can be considered as a spring which has spring constant $k$. Then the beam obeys Hook's law

\begin{equation}
\label{eqn:Hooks-law}
F = - k z.
\end{equation}

Using result from Equation (\ref{eqn:deflection-x-I-not-constant-2}) spring constant of lup-like cantilever beam tipped in $x = c_2$ can be found

\begin{eqnarray}
\nonumber
k = -\left(\frac{E H^3 B_0}{4 L^3}\right) \left\{ \frac{9 B_0^2}{8 L^3}  \left[ (1 - \alpha) L + \frac12 B_t \right] + \frac{3 B_t^2 B_0}{16 L^3} \left[ \sin^{-1} \left( \frac{B_0}{B_t} \right) - \frac{\pi}{2} \right] \right. \\
\label{eqn:spring-constant}
\left. + \frac{3 \alpha B_t}{2 L^2} \left[ (2 - \alpha)L - \frac12 B_t \right] - 3 \left( \alpha - \alpha^2 + \frac13 \alpha^3 \right) \right\}^{-1}.
\end{eqnarray}

The term in first $()$ is the spring constant for cantilever beam with constant rectangular cross section \cite{Cleveland_1993}. Or alternatively, Equation (\ref{eqn:spring-constant}) can be written in form of

\begin{equation}
\label{eqn:spring-constant-general}
k = k_\Box \beta,
\end{equation}

where $k_\Box$ is spring constant for cantilever beam with constant rectangular cross section and $\beta$ is correction factor for other form due to geometry difference

\begin{eqnarray}
\label{eqn:spring-constant-rectangular-beam}
k_\Box = \frac{E H^3 B_0}{4 L^3}, \\
\nonumber
\\
\nonumber
\beta = -\left\{ \frac{9 B_0^2}{8 L^3}  \left[ (1 - \alpha) L + \frac12 B_t \right] + \frac{3 B_t^2 B_0}{16 L^3} \left[ \sin^{-1} \left( \frac{B_0}{B_t} \right) - \frac{\pi}{2} \right] \right. \\
\label{eqn:spring-constant-beta}
\left. + \frac{3 \alpha B_t}{2 L^2} \left[ (2 - \alpha)L - \frac12 B_t \right] - 3 \left( \alpha - \alpha^2 + \frac13 \alpha^3 \right) \right\}^{-1}.
\end{eqnarray}

This factor can also put in the frequency instead in the spring constant \cite{Hoffmann_2000}. From this spring-mass system, where not all mass of the cantilever beam contributes to the oscillation (only effective mass $m^*$ instead of the whole mass $m$), the natural frequency can be found

\begin{equation}
\label{eqn:spring-natural-frequency}
\omega = \sqrt{\frac{k}{m^*}}
\end{equation}

or explicitly

\begin{equation}
\label{eqn:spring-natural-frequency-explicit}
\omega = \sqrt{\frac{E H^3 B_0 \beta}{4 L^3 m^*}}.
\end{equation}

\section{Effective mass of the vibration}

Equation of motion of a uniform beam, by neglecting shear deformation and rotary inertia, will lead to frequency equation \cite{Wu_1990}

\begin{equation}
\label{eqn:beam-frequency-equation}
1 + \cos \eta_n L + \cosh \eta_n L = 0, ~~ n = 1, 2, ..,
\end{equation}

whose solutions are related to the natural frequency of the beam vibration (with constant rectangular cross section)

\begin{equation}
\label{eqn:beam-natural-frequency}
\omega_n = \eta_n^2 \sqrt{\frac{EIL}{m}} = \eta_n^2 \sqrt{\frac{E H^3 B L}{12 m}}.
\end{equation}

For the lowest vibration frequency ($n = 1$) solution of Equation (\ref{eqn:beam-frequency-equation}) is about $1.875 / L$. Then using Equations (\ref{eqn:spring-natural-frequency-explicit}) and (\ref{eqn:beam-natural-frequency}) and the solution of frequency equation

\begin{equation}
\label{eqn:effective-mass}
\begin{array}{rcl}
\omega^2 & = & \omega_1^2 \\
&&\\
\displaystyle \frac{E H^3 B_0 \beta}{4 L^3 m^*} & = & \displaystyle \left(\frac{1.875}{L}\right)^4 ~ \frac{E H^3 B L}{12 m} \\
&&\\
\displaystyle\frac{\beta}{m^*} & = & \displaystyle \frac{1.875^4}{3m} \\
&&\\
m^* & \approx & 0.243 \beta ~m.
\end{array}
\end{equation}

For common cantiler beam with constant rectangular cross section value of $\beta$ is 1 \cite{Rabe_1996, Cleveland_1993}. It can be said that $\beta$ shows the contribution of mass from the circular head of the cantiler beam.

\section{Influence of $B_0$, $B_t$, and $L$}

\begin{figure}[h]
\center
\includegraphics[width=10cm]{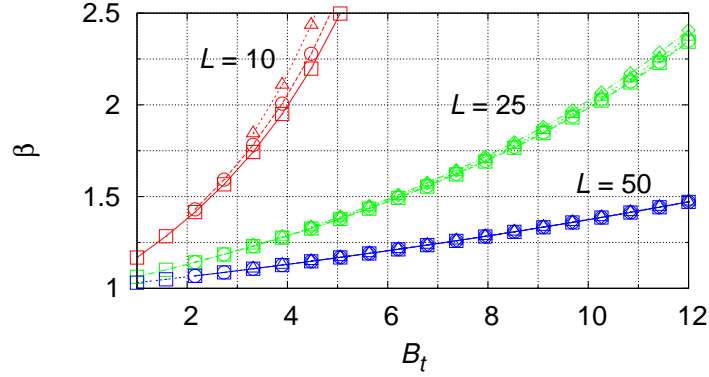}
\caption{Plot of $\beta$ as function of $B_t$ for $B_0$: 1 ($\Box$), 2 ($\Circle$), 3 ($\triangle$), and 4 ($\diamond$).}
\label{fig:beta-function-B0-Bt}
\end{figure}

With a certain beam length $L$ four values of $B_0$, which are 1, 2, 3, and 4, are used to plot $\beta$ against $B_t$ as illustrated in Figure \ref{fig:beta-function-B0-Bt}.

It can be seen from Figure \ref{fig:beta-function-B0-Bt} that $B_0$ does not play a significant role. All curves with different values of $B_0$ seem to coincide. But values of $B_t$ does change value of $\beta$ significantly but still in the same order. The curves are also groupped with the same value of $L$, which means it has also a strong influence to $\beta$. Higher value of $L$ will coincide different $B_0$ better than lower one for  the same range of $B_t$.

\section{Conclusion}

Derivation of mass, area of moment inertia, spring constant, and effective mass for lup-like cantiler beam are already presented in this work. A parameter $\beta$ is also defined, which relates efective mass for constant rectangular cross section cantiler beam to the discussed lup-like cantiler beam. Width of the beam $B_0$ does not change significantly value of $\beta$ but width of the head $B_t$ and beam length $L$ do, which has been showed graphically.

\bibliographystyle{unsrt}
\bibliography{refs}

\end{document}